# Dark Energy Stars, AdS/CFT, and High T$_c$

G. Chapline
Lawrence Livermore National Laboratory

As developed previously the physical behavior of space-time near to an event horizon can be surmised from the behavior of a superfluid near to a critical point where the speed of sound vanishes. In this talk we note that this critical behavior can be given a geometrical interpretation by regarding the tuning parameter as a radial coordinate for anti-de Sitter space. We also draw attention to a mathematical model for a layered conductor with strong spin-orbit interactions which provides a physical explanation for why the behavior of space-time near an event horizon is related to the behavior of high Tc superconductors.

## 1. Introduction

Recently much attention has been focussed on the fact that a mathematical correspondence between conformal field theories and string theory in an anti-de Sitter background (AdS/CFT) has allowed one to obtain estimates for quantum critical transport coefficients that agree with experiments [1]. Part of the excitement surrounding these results derives from the expectation that this "AdS/CFT correspondence" may also be interpreted as a theory of how classical space-time emerges from an underlying quantum theory. However, the AdS/CFT correspondence is a purely mathematical result which begs the question whether there actually is a connection between the quantum many body systems and the microscopic physics of space-time. On the other hand a connection between the behavior of condensed matter systems near to a quantum critical point and the quantum behavior of space-time near to an event horizon does appear in the theory of dark energy stars [2]. Moreover, this connection mirrors the AdS/CFT correspondence in a remarkable way.

The picture of gravitational collapse provided by classical general relativity (GR) cannot be completely correct because it conflicts with ordinary quantum mechanics during the final stages of collapse. As an alternative it has been suggested [3,4] that the interior space-time of compact astrophysical objects is a macroscopic quantum state. This assumption implies that during the final stages of the gravitational collapse the mass-energy of the matter in the collapsing object gets converted into vacuum energy. The name "*dark energy star*" has been introduced to describe the compact object formed at the endpoint of the collapse [1].

In 2000 R. Laughlin and the author realized [5] that the surface of a dark energy star corresponds to a quantum critical phase transition of space-time vacuum state analogous to the quantum critical phase transitions that have been observed in many kinds of condensed matter systems at low temperatures. The new picture that emerges for compact astrophysical objects is that there is no singularity in the interior, but the interior vacuum energy is much larger than the cosmological vacuum energy. The time dilation factor for the interior metric is positive, but becomes small as one approaches the surface. The event horizon predicted by GR is replaced by a thin surface layer where one needs new physics.

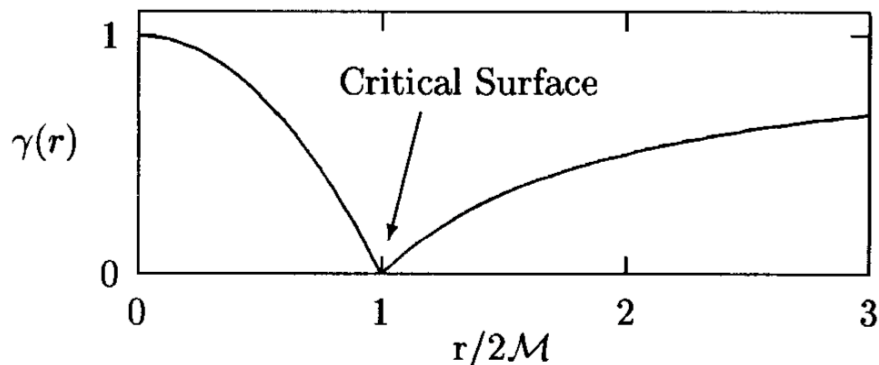

Fig. 1 Time dilation factor inside and outside a dark energy star

## 2. Warped space-time description of the critical surface layer

Two aspects of the critical behavior of superfluids are of particular interest in connection with understanding the behavior of space-time near to an event horizon. Firstly, as one approaches a critical surface the fact that the superfluid actually consists of microscopic particles will become apparent. Indeed it follows from the non-linear Schrodinger equation for a superfluid that the dispersion relation for small amplitude waves approaching a critical surface in the superfluid will has the form:

$$\hbar\omega_q = \sqrt{\left(\hbar c_s q\right)^2 + \left(\frac{\hbar^2 q^2}{2m}\right)^2}\,, \qquad (1)$$

where $c_s$ is the speed of sound.. In the case of a dark energy star $c_s = \mathrm{c}(z/2R_g)$, where $R_g = 2GM/c^2$ is the radius of the dark energy star with mass $M$ and z is the coordinate distance (using Schwarzschild or de Sitter coordinates) from the critical surface. It follows from Eq. (1) that within a distance

$$z^* = R_g \sqrt{\hbar\omega/2mc^2} \qquad (2)$$

the microscopic degrees freedom for the superfluid appear. In the context of space-time this means that classical general relativity will break down at a distance from the critical surface that depends on the energy of the particle. This dependence of the length scale where condensate particles appear on the energy of the particle is the origin of the "warped" space-time description of the critical surface. It follows from Eq's (1) and (2) that very near to the critical surface the typical momentum of particles is proportional to z. Measured in terms of proper distance from the critical surface the typical momentum increases exponentially. It happens that this is a defining characteristic of anti-de Sitter space with metric

$$ds^2 = \frac{z^2}{R_g^2}\left(dt^2 - dx_i dx_i\right) - \frac{R_g^2}{z^2}dz^2 \qquad (3)$$

As in the Boltzmann equation the setting for the microscopic quantum degrees of freedom is an internal flat 3-dimensions attached to ordinary space. Moreover, this description of the dark energy star mirrors the AdS/CFT correspondence in a remarkable way. The center of a dark energy star is a strongly coupled superfluid where continuum GR applies, while near the surface weakly coupled quantum degrees of freedom appear and classical GR breaks down. The weakly coupled particles associated with the microscopic degrees of freedom have a large mass, presumably on the order of the Planck mass.

A second important consequence of the breakdown of a continuum GR description near to z* is the appearance of dissipation and critical fluctuations. Far from the critical surface a superfluid has excitations resembling relativistic particles, and these relativistic particles can be expected to have point-like 4- point interactions. As a result of these 4-point interactions superfluid excitations can decay as they approach the critical surface producing a β-decay like spectrum for the decay products. In the context of astrophysics these decays may have observable consequences [6]. The point-like character of these interactions is also consistent with the general behavior of transport coefficients near to a quantum critical point; e.g. collision times on the order of $\hbar$/kT [7]. The prediction that this scaling law should hold for strongly coupled systems is regarded as one of the successes of AdS/CFT. On the other hand a detailed physical understanding of the critical surface where classical GR breaks down will require a specific model of the microscopic degrees of freedom for space-time.

## 3. Chiron description of the critical surface layer

In this talk we wish to draw attention to a mathematical model for a layered conductor with strong spin orbit interactions, which seems to provide both a remarkably good description of the behavior of high Tc superconductors near to the superconducting transition. In this model the 3-dimensional lattice of real materials is replaced by a stack of 2-dimensional planes [8]. The charge carriers in these 2-dimensional planes are quasi-localized solitons, which we have called "chirons" because of the key role played by chiral currents. Soon after the chiron model was put forward as a theory of high Tc superconductivity it was realized that this theory might also be interpreted as a quantum theory of space-time [9,10]. In this theory the vacuum of space-time is pictured as arising - as in the case of high Tc superconductivity - from a KT-like condensation of self-dual and anti-self-dual chirons. The resulting theory of space-time resembles the classical "ambi-twister" construction for Einstein spaces.

The chiron theory is defined by the N →∞ limit of the 2-dimensional non-linear Schrodinger equation with SU(N) Chern-Simons gauge potentials :

$$i\hbar \frac{\partial \Phi}{\partial t} = -\frac{1}{2m}D^2\Phi + e[A_0, \Phi] - g[[\Phi^*, \Phi], \Phi] \ , \qquad (4)$$

where the wavefunction $\Phi$ and gauge potentials $A_0$ and $A_i$ are N x N SU(N) matrices and $\mathbf{D} \equiv \nabla - i(e/\hbar c)[A,$ . The gauge fields corresponding to $A_0$ , and $A_i$ do not satisfy Maxwell's equations, but instead are determined self-consistently from Eq (1) and the constitutive Chern-Simons equations

$$J^\mu = \frac{\kappa}{2}\varepsilon^{\mu\alpha\beta}F_{\alpha\beta} \qquad (5)$$

relating the gauge fields to the charge densities and electric currents in the various layers. $\kappa$ is a dimensionless constant that measures the strength of the spin orbit interaction. It was shown by Grossman [11] that exact analytic time independent solutions of Eq.s (4-5) can be found for any N. In ref [8] it was shown in that in the limit N $\rightarrow \infty$ these analytic solutions take a particularly simple form such that the effective action for a gas of chirons has the form:

$$Z_c = \exp - \pi K \left[ \sum_{i \neq j} m_i m_j \ln \frac{R_{ij}}{|z_i - z_j|} \right], \qquad (6)$$

where $R_{ij}^2 = (u_i - u_j)^2 + 4(z_i - z_j)(\bar{z}_i - \bar{z}_j)$, $u_j$ is the height of the $j$-th layer, $z_j$ is the position within the jth layer, and $m_j = \pm 1$ for self-dual or anti-self-dual chirons. It can be shown that an effective interaction of the form (6) leads to a Kosterlitz-Thouless-like transition. This transition differs from the usual KT transition in that the effective interaction between chirons becomes logarithmic when the average separation between chirons is less than the interlayer spacing. The predicted transition temperature in the case of the high $T_c$ superconductor LSCO is in remarkably good agreement with what is observed in the under-doped region [12]. It turns out that exact solutions to the non-linear Schrodinger eq. (4) can also be found for the case of an external magnetic field [13,14]. Recently these exact solutions have been used to explain the unusual Nernst effect observed in high $T_c$ superconductors above the superconducting transition temperature [15]. It is of course tempting to regard the chiron picture of high $T_c$ superconductors above the transition temperature as a condensed matter model for the D-brane construction of anti-de Sitter space [16].

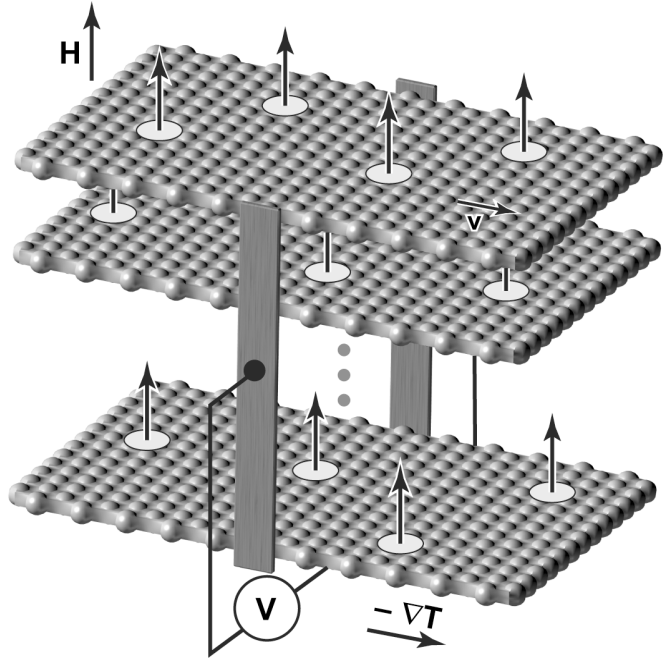

Fig. 2 Chiron theory of the Nernst effect in high $T_c$ superconductor

Acknowledgments
This work was performed under the auspices of the U.S. Department of Energy by Lawrence Livermore National Laboratory under Contract DE-AC52-07NA27344.

**References**

1. J. Zaanen, Nature **448**, 1000 (2007).
2. G. Chapline, *Proc Texas Conference on Relativistic Astrophysics, Stanford, CA 12/12-17/04* (SLAC 2005).
3. G. Chapline in *Foundations of Quantum Mechanics (*Singapore: World Scientific, 1992) pp. 255-260.
4. P. Mazur, hep-th/9701011 (1997).
5. G. Chapline, E. Hohlfeld, R. Laughlin, R. & D. Santiago, Phil. Mag. **B 81**, 235 (2001).
6. J. Barbieri and G. Chapline, astro-ph/0806.1550.
7. V. Aji and C. Varma, Phys. Rev. Lett. **99**, 067003 (2007)
8. G. Chapline and K. Yamagishi, Phys. Rev. Lett., **66**, 3046 (1991).
9. B. Grossman, Phys. Rev. Lett. **65,** 3230 (1991).
10. G. Chapline, Mod. Phys. Lett. **A14**, 2169 (1999).
11. G. Chapline, in *Proc. XXI Int. Conf. on Differential Geometric Methods in Theoretical Physics*, ed. C. N. Yang et. al. (Singapore: World Scientific 1993).
12. G. Chapline, Phil Mag **88,** 1227 (2008)
13. Z. Ezawa, M. Hotta, A. Iwazaki, Phys. Rev. **D 44**, 452 (1991).
14. R. Jackiw and S. Y. Pi, Phys. Rev. Lett. **67,** 415 (1991).
15. G. Chapline, to be published
16. J. Maldacena, Int. J, Theor. Phys. **38**, 105(1998).